\documentclass[twocolumn,prc,aps,showpacs,superscriptaddress,amsmath,amssymb,floatfix,nofootinbib,superscriptaddress]{revtex4}

\usepackage{amsmath}
\usepackage{bm}
\usepackage{dcolumn}
\usepackage{subfigure}
\usepackage{amssymb}
\usepackage{xspace}
\usepackage{graphicx}
\usepackage{epsfig}

\newcommand{\etal}{{\it et al.\/ }}

\def\be{\begin{equation}}
\def\ee{\end{equation}}
\def\bea{\begin{eqnarray}}
\def\eea{\end{eqnarray}}

\newcommand{\qsq}{\mbox{\ensuremath{Q^2}}}

\newcommand{\gevsq}{\mbox{\ensuremath{\text{GeV}\text{}^2}}}
\def\fpiq2{$F_{\pi}(Q^2)$}

\usepackage{amssymb}
\usepackage{amsmath}
\usepackage{graphicx} 
\usepackage{epsfig}   
\usepackage{dcolumn}  
\usepackage{rotating} 
\usepackage[dvips]{color}

\usepackage{float}    
\floatstyle{boxed}

\usepackage{cases}
\usepackage{hyperref}
\usepackage{breakurl}
\usepackage{color}

\begin{document}

\vspace{1.2in} NT@UW-14-05

\title{Pion transverse charge density and the edge of hadrons}

\newcommand*{\CUA}{The Catholic University of America, Washington, D.C. 20064, USA}
\newcommand*{\JLAB}{Thomas Jefferson National Accelerator Facility, Newport News, VA 23606, USA}
\newcommand*{\UW}{University of Washington, Seattle, Washington 98195, USA}

\author{Marco Carmignotto} 
\affiliation{\CUA}

\author{Tanja Horn} 
\affiliation{\CUA}

\author{Gerald A. Miller}
\affiliation{\UW}

\newpage
\date{September 18, 2014}
\newpage


\begin{abstract}
We use the world data on the pion form factor for space-like kinematics and a technique previously used to extract the proton transverse densities to extract the transverse pion charge density and its uncertainty due the incomplete knowledge of the pion form factor at large values of $Q^2$ and the experimental uncertainties. The pion charge density at small values of impact parameter $b<$~0.1~fm is dominated by this incompleteness error while the range between 0.1-0.3~fm is relatively well constrained. A comparison of pion and proton transverse charge densities shows  that the pion is denser than the proton for values of $b<$~0.2~fm. The pion and proton transverse charge densities seem to be the same for values of $b$=0.3-0.6~fm. Future data from Thomas Jefferson National Accelerator Facility (JLab) 12~GeV and the Electron-Ion Collider (EIC) will increase the dynamic extent of the form factor data to higher values of $Q^2$ and thus reduce the uncertainties in the extracted  pion transverse charge density.
\end{abstract}

\maketitle


\section{Introduction}

Measurements of form factors play an important role in our understanding of the structure and interactions of hadrons based on the principles of quantum chromodynamics (QCD). One of the simplest hadronic systems available for study is the pion, whose valence structure is a bound state of a quark and an antiquark. Its elastic electromagnetic structure is parameterized by a single form factor $F_\pi(Q^2)$. Calculations  of the pion charge form factor have been used to investigate  the transition from the low-momentum transfer confinement region to the regime where perturbative methods are applicable\cite{Lepage:1980fj,Efremov:1979qk}.  There is a long history of experimental measurements~\cite{ady77}-\cite{horn08}. In particular, $F_\pi(Q^2)$ has been measured at space-like momentum transfers through pion-electron scattering and pion electroproduction on the nucleon with high precision up to $Q^2$=2.5~GeV$^2$, and new measurements are planned with the 12~GeV era at the Thomas Jefferson National Accelerator Facility (JLab)~\cite{E12-06-101, E12-07-105} and envisioned for a future Electron-Ion Collider (EIC)~\cite{huber09}. 

The concept of transverse charge densities~\cite{Soper77} has emerged recently~\cite{Miller07, Burkardt03} as a framework providing an interpretation of electromagnetic form factors in terms of the physical charge and magnetization densities. It has been explored in a number of recent works~\cite{Carlson:2007xd,Miller09,Miller10, Strikman10, Miller11, Venkat11}. These transverse densities are obtained as two-dimensional Fourier transforms of elastic form factors and describe the density of charge and magnetization in the plane transverse to the propagation direction of a fast moving nucleon. They are related to the generalized parton distributions (GPDs)~\cite{goeke01, diehl03, belitsky05}, which are expected to provide a universal (process-independent) description of the nucleon, and simultaneously encode information on parton distributions and correlations in both momentum (in the longitudinal direction) and coordinate (in the transverse direction) spaces. 

There have been two previous analyses of the pion transverse charge density~\cite{Miller09,Miller11}. In the first a wide range of models was used. No estimate of the uncertainty caused by incomplete kinematic knowledge of the form factor was made. The second was based on data taken in the time-like region and extended to the space-like region through the use of dispersion relations and models needed to obtain the separate real and imaginary parts of the observable quantity $|F_\pi(Q^2)|^2$. The present paper is aimed at avoiding models and determining the impact of potential new experiments.
 
In particular, the goal of the present analysis is to evaluate the world's data on the space-like pion form factor, to extract the corresponding pion transverse charge density within current uncertainties, and to estimate the influence of the planned experiments on the pion transverse charge density. Examining the current data requires forming a superset with a single global uncertainty, taking into account the individual uncertainties and the differences in the form factor extraction method. This is done in Sec. II. We use the finite radius approximation technique applied to analyze the proton form factor data described in Ref.~\cite{Venkat11} to estimate the uncertainty due to the limited kinematic coverage of the currently available data in Sec. III. Results for the pion transverse charge density are presented in Sec. IV. An interesting application of transverse charge densities is the analysis of the spatial structure of the proton's pion cloud. Recent work~\cite{Strikman10} found that the non-chiral core of the charge density is dominant  up to rather large distances $\sim$ 2~fm implying a large proton core. The proton and pion transverse charge density are compared in Sec. V, and the impact of future experiments is assessed in Sec. VI. Our analysis is consistent with the general trends of the pion charge density reported by the authors of Ref.~\cite{Miller09}, the present analysis is of higher precision and more extensive.

\section{\label{sec:fpi_data}Extraction of the pion form factor from world data}

The pion's elastic electromagnetic structure is parameterized by a single form factor $F_\pi(Q^2)$, which depends on $Q^2=-q^2$, where $q^2$ is the four-momentum squared of the virtual photon. $F_\pi(Q^2)$ is well determined up to values of $Q^2$ of 0.28 GeV$^2$ by elastic $\pi-e$ scattering~\cite{ady77,dal82,ame86,ame84}, from which the charge mean radius of the pion has been extracted. Determining $F_\pi(Q^2)$ at larger values of $Q^2$ requires the use of pion electroproduction from a nucleon target. The longitudinal part of the cross section for pion electroproduction $\sigma_L$ contains the pion exchange process, in which the virtual photon couples to a virtual pion inside the nucleon. This process is expected to dominate at small values of the Mandelstam variable $-t$, thus allowing for the determination of $F_\pi(Q^2)$. A comprehensive review on the extraction of $F_\pi(Q^2)$ from pion electroproduction data can be found in Refs.~\cite{blok08,huber08}.

Pion electroproduction data have previously been obtained for values of \qsq\ of 0.18 to 9.8~\gevsq\ at the Cambridge Electron Accelerator (CEA), at Cornell~\cite{beb76, beb78} and at the Deutsches Elektronen-Synchrotron (DESY)~\cite{bra77, ack78}. Most of the high $Q^2$ data have come from experiments at Cornell. In these experiments, $F_\pi(Q^2)$ was extracted from the longitudinal cross sections, which were isolated by subtracting a model of the transverse contribution from the unseparated cross sections. Pion electroproduction data were also obtained at DESY~\cite{ack78,bra76,bra77} for values of $Q^2$ of 0.35 and 0.7~GeV$^2$, and longitudinal (\textit{L}) and transverse (\textit{T}) cross sections were extracted using the Rosenbluth \textit{L}/\textit{T} separation method. With the availability of the high-intensity, continuous electron beams and well-understood magnetic spectrometers at JLab it became possible to determine \textit{L}/\textit{T} separated cross sections with high precision, and thus to study the pion form factor in the regime of \qsq=0.5-3.0~\gevsq~\cite{vol01, tad06, horn06, horn08, horn12}.

The pion form factor has been compared to different empirical fits and model calculations based on perturbative quantum chromodynamics (pQCD), lattice QCD, dispersion relations with QCD constraint, QCD sum rules, Bethe-Salpeter equation, local quark-hadron duality, constituent quark model, holographic QCD, and so on in Ref.~\cite{huber08}. See also Ref.~\cite{masjuan08} for details on a method using rational approximants. A new method has recently been developed to calculate $F_\pi(Q^2)$ on the entire region of $Q^2$ using the Dyson-Schwinger equation framework~\cite{Chang13}. The results are in very good agreement with the world $F_\pi(Q^2)$ data. Many models in the literature approach the monopole form $F_\pi^{monopole}(Q^2)~=~1/\left(1+Q^2r_\pi^2/6\right)$ for large values of $Q^2$, where $r_\pi$ is a measure of  the slope of $F_\pi(Q^2)$ at $Q^2=0$ via $r_\pi^2\equiv -6 F_\pi'(Q^2=0)$ is denoted as the radius of the pion, and  is often chosen as the inverse of the mass of the $\rho$ meson.

Our analysis of the uncertainty due to lack of $F_\pi(Q^2)$ data at values of $Q^2>$~9.8~GeV$^2$ requires the use of an upper bound and a lower bound on $F_\pi(Q^2)$ in the region where it is not measured. An upper bound~\cite{blok08,huber08} for the pion form factor is given by the monopole form, so we use this form with $r_\pi=0.672\pm0.008$~fm~\cite{beringer12} to provide the upper bound in our analysis. Our lower bound is chosen to be a light front constituent quark model that does not converge to the monopole asymptotically yet still describes the data well. There are many models available in this category, which typically differ in the treatment of the quark wave functions of relativistic effects. The model of Ref.~\cite{Hwang01} provides a relativistic treatment of quarks spins and center of mass motion. It uses a power-law wave function with parameters determined from experimental data on the charged pion decay constant, the neutral pion two-photon decay width, and the charged pion electromagnetic radius. This model is in very good agreement with the world $F_\pi(Q^2)$ data.

\begin{figure}[H]
\begin{center}
\epsfxsize=3.0in
\epsfysize=3.0in
\epsffile{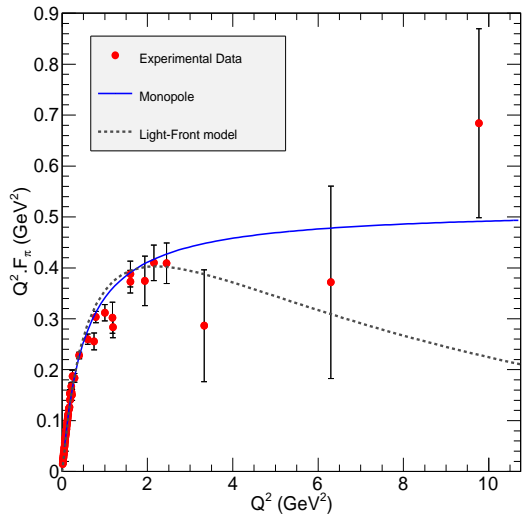}
\caption{\label{fig:global:data} (Color online) Red points are global data of the pion form factor from pion-electron scattering and pion electroproduction on the nucleon. The error bars represent the statistical and systematic uncertainties of the individual measurements. The monopole empirical form (blue line) and the Light-Front model (dashed-gray line) both describe the existing $F_\pi(Q^2)$ data well. These models are constrained by the existing data, but differ in their asymptotic behavior.}
\end{center}
\end{figure}

Figure~\ref{fig:global:data} shows the world data for the pion form factor together with the results of the empirical monopole form and a light front model (LF) calculation based on that found in Ref.~\cite{Hwang01}. Both the monopole and the LF models are in very good agreement with the data up to values of $Q^2~\approx$~2.5~GeV$^2$. Above that, $Q^2F_\pi^{monopole}(Q^2)$ and $Q^2F_\pi^{LF}(Q^2)$ deviate from each other. $Q^2F_\pi^{monopole}(Q^2)$ tends to a constant value while $Q^2F_\pi^{LF}(Q^2)$ decreases as $Q^2\rightarrow\infty$. All other models of the pion form factor fall between these two curves. No distinction can be made between the models based on the current data due to their large uncertainties in particular at values of $Q^2$ between 3~and 10~GeV$^2$.

Elastic pion-electron scattering has been measured up to 0.28~GeV$^2$ and the pion form factor has been extracted with high precision up to $Q^2$=2.5~GeV$^2$ from pion electroproduction data. The two main sources of uncertainty in the extraction of the transverse densities are the published experimental uncertainties from the measurements of the pion form factor and uncertainties due to the lack of form factor data at values of $Q^2>$~9.8~GeV$^2$. 

Here we present a new global analysis of the world $F_\pi(Q^2)$ data to obtain a parametrization of the pion form factor that enables the uncertainties in the transverse charge density to be determined. The current $F_\pi(Q^2)$ data show a systematic departure from the monopole curve above $Q^2~\sim$~1.5~GeV$^2$. In our analysis we thus use a three parameter empirical fit form as follows:

\begin{equation}
F_\pi\left(Q^2\right) = A\cdot\frac{1}{\left(1+B\cdot{}Q^2\right)}+\left(1-A\right)\cdot\frac{1}{\left(1+C\cdot{}Q^2\right)^2} .
\label{eq:global:empirical}
\end{equation}

\begin{figure}[H]
\begin{center}
\epsfxsize=3.0in
\epsfysize=3.0in
\epsffile{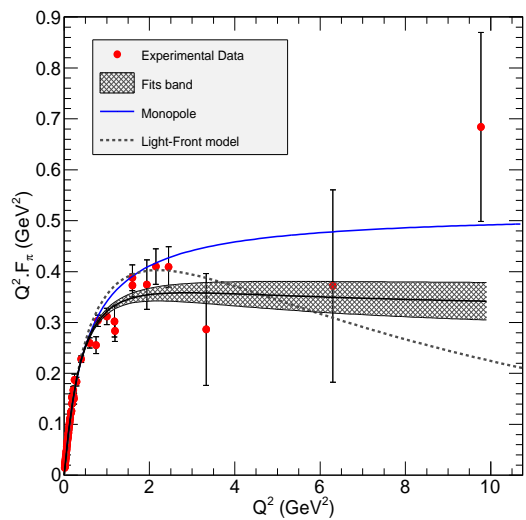}
\caption{\label{fig:global:multifit}(Color online) Empirical fit to the experimental $F_\pi(Q^2)$ data (black/hatched band) used to evaluate the pion transverse charge density. The band represents the systematic uncertainties due to combining the different measurements. The error bars represent the statistical and systematic uncertainties of the individual measurements. The curves are the same as shown in Fig.~\ref{fig:global:data}.}
\end{center}
\end{figure}

The parameter \textit{A} denotes the fractional contribution of the two terms to the overall fit. Equation~\ref{eq:global:empirical} imposes the normalization condition $F_\pi(Q^2=0)=1$. We vary all of the values of the parameters \textit{A}, \textit{B} and \textit{C} simultaneously to obtain the present fit. Note that the slope of $F_\pi(Q^2)$ at $Q^2\rightarrow$0~GeV$^2$ is constrained by the world data set for  low values of $Q^2$~\cite{beringer12} and our fit incorporates this information. Curves with the form of Eq.~\ref{eq:global:empirical} were fitted to the data. For each fit, the experimental points were randomly recreated following a Gaussian distribution around their central values. The results of these fits are shown in the black/hatched band in Fig.~\ref{fig:global:multifit}. We find the best coefficients for these fits to be $A=0.384\pm0.071$, $B=1.203\pm0.101$~GeV$^{-2}$ and $C=1.054\pm0.080$~GeV$^{-2}$ with $\chi^2$=1.64, corresponding to a probability of 99\%. Using these coefficients we extract a value of $r_\pi$ of 0.641~$\pm$~0.025~fm, which is consistent with the value extracted from the world data $0.672\pm0.008$~fm~\cite{beringer12}. 

The dominance of the first term over the second term in the present fit differs from the result of the authors of Ref.~\cite{huber08}, who found the first dominant. The constraints on the fit in Ref.~\cite{huber08} are different from our present fit in that their values of \textit{B} and \textit{C} were kept fixed and only the fractional contribution \textit{A} was fitted. Furthermore, our present fit included additional data points up to $Q^2=$9.8~GeV$^2$. The impact of the additional higher $Q^2$ data points on the present fit is small due to their large experimental uncertainties; the fit parameters change by less than 0.5\%. However, including these points here despite their large uncertainties is important for the truncation of the series expansion in Eq.~\ref{eq:theory:bigEq}, and the resulting incompleteness error. This error results from the region in $Q^2$ where no measurements exist at all. As Fig.~\ref{fig:incompletenessError} shows, the incompleteness error dominates over the experimental error.

\section{Extraction of the pion transverse charge density}

We apply the method of the authors of Ref.~\cite{Venkat11} to studying the pion. In particular, the pion transverse charge density $\rho_{\pi}(b)$ is the matrix element of the LF density operator integrated over longitudinal distance~\cite{Miller10} and is given by the two-dimensional Fourier transform of the space-like pion form factor $F_\pi(Q^2)$

\begin{equation}
\rho_{\pi}(b) = \frac{1}{(2\pi)^2}\int{}\mathrm{d}^2q e^{-i\vec{q}\cdot\vec{b}}F_\pi(Q^2) ,
\label{eq:theory:fourierA}
\end{equation}

where $\vec{q}\,^2=Q^2$. The transverse density $\rho_{\pi}(b)$ denotes the probability that a charge is located at a transverse distance $b$ from the transverse center of momentum with normalization condition, $\int{}\mathrm{d}^2b \rho_{\pi}(b)$=1. If we consider the azimuthal symmetry of $\rho_{\pi}$, Eq.~\ref{eq:theory:fourierA} reduces to a one-dimensional integral

\begin{equation}
\rho_{\pi}(b) = \frac{1}{2\pi}\int_0^\infty Q\mathrm{d}Q J_0\left(Qb\right)F_\pi\left(Q^2\right) .
\label{eq:theory:fourierB}
\end{equation}

Intuitively we expect the charge of the pion to be localized within a volume of radius $R$. This assumption is called the {\it finite radius approximation}~\cite{Venkat11} and we use it to simplify Eq.~\ref{eq:theory:fourierB}. For values of $b$ less than the chosen distance parameter $R$, the function $\rho_{\pi}(b)$ can be expanded in a series of the Bessel function $J_0$ as

\begin{equation}
\rho_{\pi}(b) = \sum_{n=1}^\infty c_nJ_0\left(X_n\frac{b}{R}\right) ,
\label{eq:theory:series}
\end{equation}

where $X_n$ is the {\it n}-th zero of $J_0$ and $c_n$, as obtained from the orthogonality of the Bessel functions over the range $0\le b\le R$ is given by the expression

\begin{equation}
c_n = \frac{1}{2\pi}\frac{2}{R^2\left(J_1\left(X_n\right)\right)^2}F_\pi\left(Q_n^2\right),
\label{eq:theory:coeffs}
\end{equation}

with $Q_n$  defined as

\begin{equation}
Q_n \equiv \frac{X_n}{R} .
\label{eq:theory:Qn}
\end{equation}

Combining Eqs.~\ref{eq:theory:series} and \ref{eq:theory:coeffs} yields the following expression for $\rho_{\pi}(b)$:

\begin{equation}
\rho_{\pi}(b) = \frac{1}{\pi{}R^2}\sum_{n=1}^\infty F_\pi\left(Q_n^2\right)\frac{J_0\left(X_n\frac{b}{R}\right)}{\left(J_1\left(X_n\right)\right)^2} .
\label{eq:theory:bigEq}
\end{equation}

This expansion provides the transverse density for values of $b<R$ for measurements of the pion form factor up to momentum transfers of $Q^2_{max}$.

The extraction of the pion transverse density requires the experimental value of $F_\pi(Q^2)$ obtained from the fits shown in Fig.~\ref{fig:global:multifit} as input. The uncertainty on the extraction thus also depends on the experimental uncertainties. The total uncertainty on $\rho_{\pi}(b)$ has two main sources: (1) experimental uncertainties on the individual measurements and combining data from different experiments in the region where data exist for $Q^2\le Q^2_{max}=10 $ GeV$^2$ and (2) uncertainties due to the lack of data in the region $Q^2>Q^2_{max}$, where no measurements exist. The experimental uncertainties are taken into account directly in the  through Eq.~\ref{eq:theory:bigEq}. However, uncertainty due to lack of data for values of $Q^2>Q^2_{max}$ must also be estimated. Both sources of uncertainty are discussed next.

\subsection{Experimental Uncertainty}

\begin{figure}[H]
\begin{center}
\epsfxsize=3.0in
\epsfysize=3.0in
\epsffile{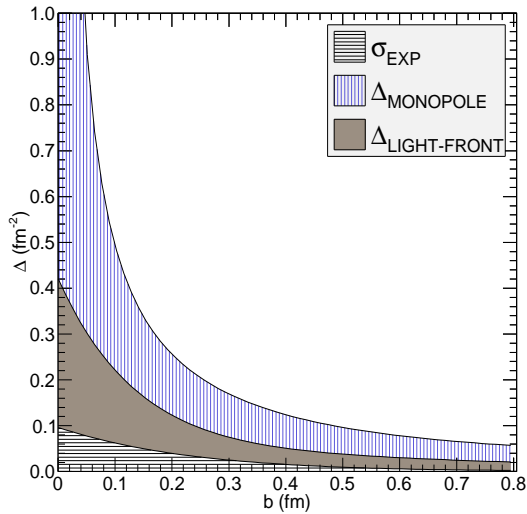}
\caption{\label{fig:incompletenessError} (Color online) Uncertainties of the pion transverse charge density due to $F_\pi(Q^2)$ experimental data uncertainties $\sigma_\text{EXP}$ (black/hatched band) and the incompleteness error, when considering the monopole model (blue band) or the LF model (gray band). The incompleteness error is estimated using the difference between two models that describe the existing data well, but have different asymptotic behavior: $\Delta_\text{MONOPOLE}$ and $\Delta_\text{LIGHT-FRONT}$. $\Delta_{\text{MONOPOLE}}$ goes to infinity as $b\rightarrow$0~fm due to the $1/Q^2$ asymptotic behavior of the monopole form. The total uncertainty on the pion transverse charge density is the sum of the experimental uncertainties and the incompleteness error. This provides an uncertainty band and all existing data and all other models that describe them fall in between. Future data will narrow this band as discussed in the text.
}
\end{center}
\end{figure}

The form factor $F_\pi(Q^2)$ has been measured with high precision up to $Q^2$=2.5 GeV$^2$. We also use lower precision data with large systematic uncertainties 50-70\% for values $Q^2=$3.3-9.8~GeV$^2$.  Thus we take the form factor to be a measured quantity for $Q^2<Q^2_{max} =10$~GeV$^2$. The series expansion in Eq.~\ref{eq:theory:bigEq} is truncated to values of $Q^2_{n}$ for which the pion form factor has been extracted from data. Uncertainties from these data causes uncertainties in $\rho_\pi$ via Eq.~\ref{eq:theory:bigEq}. This corresponds to an upper limit of $n_{max}=10$, when using $R=2$~fm, in Eq.~\ref{eq:theory:Qn}. The contribution of the experimental uncertainty to the pion transverse charge density is illustrated by the black/hatched area shown in Fig.~\ref{fig:incompletenessError}. We see that the largest uncertainty of $0.1 {\rm \,fm}^{-2}$ occurs at $b=0$, consistent with the uncertainty principle which relates distance to momentum.

We use $R$ for the pion smaller than the used for the proton because the the pion radius, $r_\pi=0.672\pm0.008$~fm is smaller than the corresponding value for the proton of either 0.84 or 0.87~fm~\cite{Pohl:2013yb}. Furthermore, proton form factors have been measured accurately at much higher values of $Q^2$ than for the pion. Testing the sensitivity of our results for choices of $R$ larger than 2.0~fm shows that the pion transverse charge density does not change within the stated uncertainties.

\subsection{Incompleteness Error}

Uncertainties due to lack of knowledge of the pion form factor for values of $Q^2>Q^2_{max}$, where no measurements exist, must be estimated~\cite{Venkat11}. The first step is understanding the necessary truncation. Low $Q^2$ pion form factor data are readily available and as a result the value of the transverse charge density is well defined and its uncertainty is small for large impact parameters $b$. For values of  $Q^2$  above $Q^2>$3 GeV$^2$, which corresponds to shorter distances, pion form factor data become very sparse and there are  no data available for values of $Q^2>10$~GeV$^2$.

Equation~\ref{eq:theory:bigEq} uses the finite radius approximation requiring knowledge of the form factor in the full range of $Q^2$. Since $F_\pi(Q^2)$ measurements are limited to a region $Q^2$=$Q^2_{max}$, where $Q^2_{max}$ denotes the limit of currently available pion form factor data, the series expansion has to be truncated to $Q^2_{max}$=10~GeV$^2$. The effects of this truncation in the calculation of $\rho_{\pi}(b)$ is estimated in the \textit{incompleteness error}.

The basic transverse pion densities are obtained using Eq.~\ref{eq:theory:bigEq} for values of $Q^2~<$~10~GeV$^2$ corresponding to values of $n_{max}$=10. A maximum error was estimated using two representative theoretical models with very different asymptotic behavior, which describe the existing $F_\pi(Q^2)$ data well. Out of the available models we chose the monopole (with pion RMS radius $r_\pi=0.672$~fm) and the LF model from Ref.~\cite{Hwang01}, which are both constrained by data, as an upper and lower bound, respectively. The region in between these two model predictions constitutes a band that includes the existing data and all other models of $F_\pi(Q^2)$ mentioned above in Sec.~\ref{sec:fpi_data} that describe the data. The band thus also includes the true value of $F_\pi(Q^2)$ in the region where no data exist. The incompleteness error for our two chosen models is estimated using

\begin{equation}
\Delta_{model}(b)=\left|\frac{1}{\pi{}R^2}\sum_{n=n_{max}+1}^\infty F_\pi^{model}\left(Q_n^2\right)\frac{J_0\left(X_n\frac{b}{R}\right)}{\left(J_1\left(X_n\right)\right)^2}\right|,
\label{eq:incompleteness}
\end{equation}

as a function of $b$, where $n_{max}=$10 is the last term of the transverse charge density series where $F_\pi(Q_{n_{max}}^2)$ has been measured, i.e., the tenth term of the series is evaluated at $Q_n^2$ = 9.14 GeV$^2$, below the last currently available data point of the pion form factor. The results are summarized in Fig.~\ref{fig:incompletenessError}, which  shows that the uncertainty due to  incompleteness is much larger  than that caused by uncertainty in current data. The blow up of the incompleteness error estimated using the monopole model at \textit{b}=0~fm is a consequence of its asymptotic behavior (~$\sim$ $1/Q^2$), which results in a singularity at the center of the pion~\cite{Miller09,Miller11}. Though the mean value of the pion transverse charge density is not singular at $b$=0, our results are compatible with such a singularity within the uncertainty. The incompleteness error is likely overestimated as we chose two very extreme models resulting in a very conservative incompleteness error band (given by the difference between the incompleteness error calculated using the monopole and LF models). Future high $Q^2$ data like those discussed in Sec.~\ref{sec:future_data} will significantly narrow down the error band, by constraining the models and thus reducing the incompleteness error at intermediate and small distances.

\section{Pion transverse charge density}

\newcommand {\boldlambda}{\mbox{\boldmath$\lambda$}}

\begin{figure}[H]
\begin{center}
\epsfxsize=8.6cm
\epsfysize=6.88cm
\epsffile{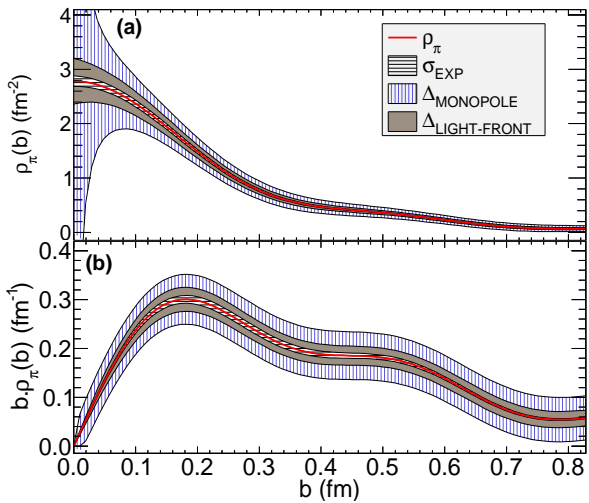}
\caption{\label{fig:DistBand} (Color online) \textbf{(a)} The pion transverse charge density (red curve) calculated from the two-dimensional (2D) Fourier transform of the pion form factor. \textbf{(b)} The pion transverse charge density multiplied by the Jacobian {\it b}. Uncertainties from experimental $F_\pi(Q^2)$ data are represented by the black/hatched band, while the incompleteness error was estimated using the monopole model (blue band) and the LF model (gray band).}
\end{center}
\end{figure}

We turn to our stated goal of using the world data on the space-like pion form factor to extract the pion transverse charge density. Figure~\ref{fig:DistBand} shows the pion charge density evaluated using the series expansion of Eq.~(\ref{eq:theory:bigEq}) with the experimental uncertainty based on our fits of $F_\pi(Q^2)$ (from Fig.~\ref{fig:global:multifit}) and with the incompleteness error estimated using the monopole and LF models as described above in Eq.~\ref{eq:incompleteness}.

As we are working in polar coordinates, the spatial transverse element of area is $d^2b=2\pi b db$, for a given impact parameter $b$. Thus Fig.~\ref{fig:DistBand}-b shows the pion transverse charge density multiplied by the Jacobian {\it b}.

For $\rho(b)$ the uncertainties due to the incompleteness error for $b>0.1$~fm are relatively small compared to the ones for the region $b<0.1$~fm. This is because pion form factor data are readily available at low values of $Q^2$ (large values of impact parameter $b$) and, as a result, $\rho(b)$ is well determined in that region. On the other hand, in the region $b<0.1$~fm the incompleteness error is very large, which is due to the lack of the pion form factor data at very large values of $Q^2$. The oscillatory behavior can be attributed to the truncation of the Bessel function series of  Eq.~\ref{eq:theory:bigEq}. However, the choice of $R$=2 fm is not physically relevant for our result. Tests of the sensitivity of our results for $R$ larger than 2.0 fm show no significant change in the pion transverse charge density within the uncertainties.  The oscillations are due to the finite range in $Q^2$ of the experimental data available for the Fourier transform. Using  values of $R$ larger than 2.0~fm increases the number of terms in the series, but  does not reduce the oscillations of the incompleteness error.

\section{Proton pion cloud and pion charge density}

\begin{figure}[H]
\begin{center}
\epsfxsize=8.6cm
\epsfysize=6.88cm
\epsffile{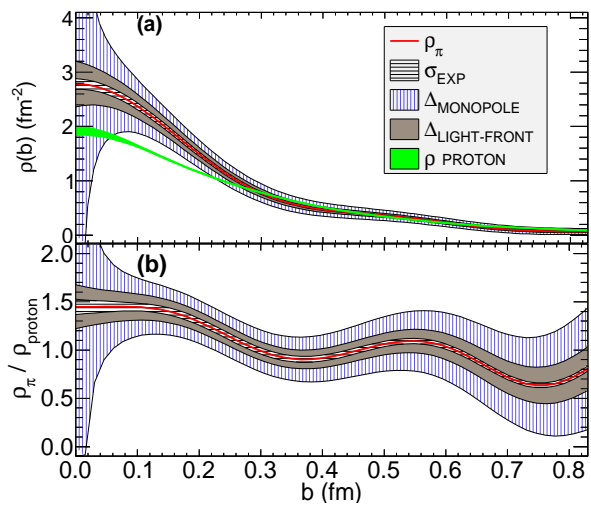}
\caption{\label{fig:proton} (Color online) \textbf{(a)} Comparison of the pion transverse charge density (red curve) to the proton transverse charge density (from Ref.~\cite{Venkat11}) shown in the green band. The uncertainties for the pion transverse charge density are as in Fig.~\ref{fig:DistBand}. The green band for the proton includes both the experimental and incompleteness error. The proton error band is smaller as compared to the pion because the proton form factor is well known over a larger range in $Q^2$. The two transverse charge density curves (red solid and green solid lines) coalesce in the region $b>$0.3~fm within the uncertainty while the pion transverse charge density appears denser than that of the proton in the region $b<$0.2~fm. \textbf{(b)} Ratio of pion to proton transverse charge densities (red solid curve). Here the error bands shown denote the uncertainty on the ratio of pion to proton charge density. The error band is dominated by the pion incompleteness error, so we keep the same coloring and shading as in panel (a) of this figure to indicate the individual uncertainty contributions.}
\end{center}
\end{figure}

\begin{figure}[H]
\begin{center}
\epsfxsize=8.6cm
\epsfysize=6.88cm
\epsffile{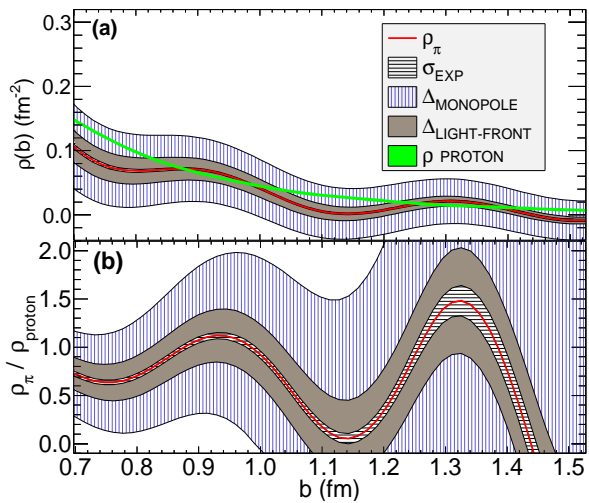}
\caption{\label{fig:proton2} (Color online) \textbf{(a)} Comparison of the pion transverse charge density to the proton transverse charge density as in Fig.~\ref{fig:proton}, showing a range of higher impact parameter \textit{b}, where the transverse charge density is expected to already be significantly smaller than at the core and to approach zero. \textbf{(b)} Ratio of pion to proton transverse charge densities. The curves and uncertainties are as in Fig.~\ref{fig:proton}.} 
\end{center}
\end{figure}

Recent work~\cite{Strikman10} explored the proton transverse charge density finding that the non-chiral core is dominant up to relatively large distances of $\sim$~2~fm. This suggests that there is a non-pionic core of the proton, as one would obtain in the constituent quark or vector meson dominance models. One does not usually think of the pion having a meson cloud since a, e.g., $\rho\pi$ component  would involve a high excitation energy. Therefore it is interesting to compare the proton and pion transverse charge densities as given by numerically stable series as in Eq.~\ref{eq:theory:bigEq}. Figures~\ref{fig:proton} and~\ref{fig:proton2} compare proton and pion transverse charge densities for different ranges of \textit{b}. Figure~\ref{fig:proton} focuses on a region in $b$ where the transverse charge density is expected to decrease from its value at the core while Fig.~\ref{fig:proton2} illustrates a region in which the transverse charge density is expected to  be significantly smaller than at the core and to approach zero.

For values of $b$ less than about 0.2~fm the transverse charge density  of the pion is larger than that of the proton. This higher density is  expected because the pion's radius $0.672$~fm is smaller than the proton's $0.84$~fm. As previously noted~\cite{Miller09,Miller11}, it is possible that the pion's transverse density is singular for small values of $b$. An interesting feature is that the curves seem to coalesce in the region $b>$~0.3~fm (at least within current uncertainties). This is not expected. A possible explanation could be obtained by regarding the pion to be a $q\bar{q}$ pair bound by a color octet exchange mechanism (proportional to $\boldlambda_i\cdot\boldlambda_j$, where the eight components of $\boldlambda_i$ are generators of SU(3) in color space) and regarding the proton as a quark-diquark~\cite{Wilson:2011aa,Cloet:2012cy} system that is also bound via a color octet exchange mechanism. Similarity in binding forces could lead to a similarity in transverse densities. The result that the pion and proton transverse densities are similar in their core may be a first experimental glimpse at the transition between proton core and meson cloud, e.g., the ``edge'' of the proton. 

The comparison between the pion and the proton transverse charge densities for \textit{b}~$\textgreater$~0.7~fm is shown in Fig.~\ref{fig:proton2}. Both transverse charge densities in this higher region of \textit{b} are significantly smaller than at the core. The uncertainties on the pion transverse charge density in this region are on the order of the charge density itself making it difficult to compare to the proton. However, the two  densities are the same, within current uncertainties. To explore any similarity of the pion and the proton transverse charge densities in this region, precision pion form factor data at higher values of $Q^2$ like those discussed in Sect.~\ref{sec:future_data} would be needed.

\section{\label{sec:future_data} Impact of future experiments}

\begin{figure}[H]
\begin{center}
\epsfxsize=3.0in
\epsfysize=3.0in
\epsffile{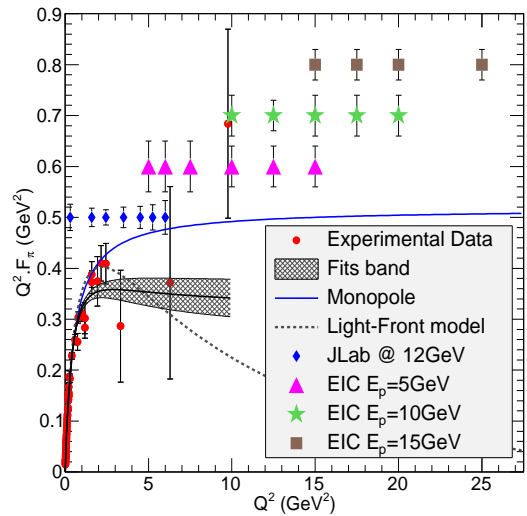}
\caption{\label{fig:FutureFF} (Color online) $F_\pi(Q^2)$ world data (red circles) and projected uncertainties for experiments at JLab at 12~GeV (blue diamonds) and those for measurements of the form factor with an EIC~\cite{huber09,E12-06-101}. The projected uncertainties at the EIC are divided into three groups depending on the energy $E_p$ of the ion beam (magenta triangles with $E_p$=5 GeV, green stars with E$_p$=10 GeV, brown squares with $E_p$=15 GeV). The error bars shown include both projected statistical and systematic uncertainties. The black/hatched band represents an empirical fit using Eq.~\ref{eq:global:empirical} taking into account present data.}
\end{center}
\end{figure}

Future data would improve the extraction of the pion transverse charge density, which in the region of $b>$ 0.3~fm would be of great interest for further studies of the ``edge'' of hadrons. Experiments at the 12~GeV JLab~\cite{E12-06-101,E12-07-105} will extend the $Q^2$ range of high precision pion form factor data to $Q^2$=~6 and 9~GeV$^2$. The envisioned EIC will further extend this reach to $Q^2$ of about 25~GeV$^2$ \cite{huber09}. This $Q^2$ region would add data into the region of interest for studying the hadron edge improving the precision of the extraction of the pion transverse charge density. The measurements would also add data into thus far unmeasured regions of small $b$. The projected uncertainties of these future experiment are shown in Fig.~\ref{fig:FutureFF} together with existing data and the monopole and LF calculations.

\begin{figure}[H]
\begin{center}
\epsfxsize=8.6cm
\epsfysize=6.88cm
\epsffile{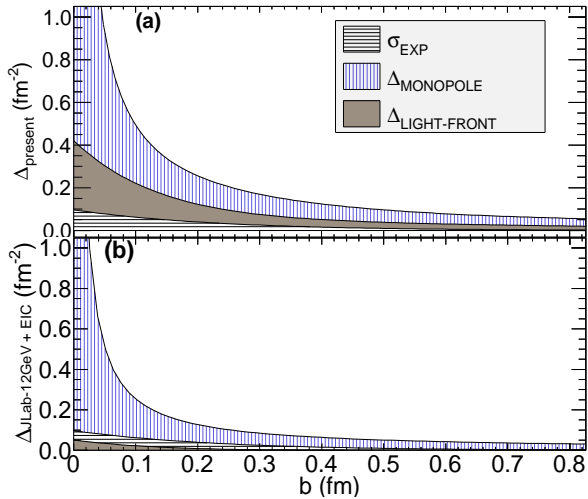}
\caption{\label{fig:FutureRho} (Color online) Experimental and incompleteness error on the pion transverse charge density considering \textbf{(a)} the existing data and \textbf{(b)} including the projected uncertainties of future $F_\pi(Q^2)$ data from 12~GeV experiments at JLab and the EIC. Uncertainties may be even lower depending on how new data constrain the existing models of the pion form factor.}
\end{center}
\end{figure}

The projected uncertainties of the new data will add sufficient precision to distinguish between theoretical models, like those mentioned in Ref.~\cite{huber08}, at values of $Q^2$ greater than 3~GeV$^2$, and thus narrow down our selection of models for estimating the incompleteness error. In our current estimation of the incompleteness error as presented in Fig.~\ref{fig:incompletenessError} we conservatively chose two models that represent extreme values of the pion form factor. The band in between the monopole and LF model curves contains any existing data and predictions from all other models that describe $F_\pi(Q^2)$. Pion form factor models are well constrained at small $Q^2$ where data are available. The model predictions begin to diverge at values of $Q^2>$3 GeV$^2$ as there are currently very few or no data available. With future data at values of $Q^2>$3 GeV$^2$ the number of models describing the data will be better constrained over a wider kinematic range, and thus the uncertainty band described by the band between the upper bound (currently using monopole) and lower bound (currently using LF model) will also become narrower. This is illustrated in Fig.~\ref{fig:FutureRho} assuming that all data from both 12~GeV JLab and the EIC are measured.

Including the projected uncertainties of the future data the precision of the pion transverse charge density would be better than 20\% for $b>0.1$~fm. This would greatly constrain the pion transverse charge density and determine the proton and pion transverse charge densities really are the same for moderate values of $b$.

\section{Summary}

In this paper we used the world data on the space-like pion form factor to extract the pion transverse charge density. The extraction method is based on that used for the proton in Ref.~\cite{Venkat11} and includes the use of Bessel series expansion and finite radius approximation to determine the impact of experimental uncertainties and the incompleteness error due to the lack of data for $Q^2~>~Q^2_{max}$.  Two theoretical models, monopole and LF, are used to estimate the incompleteness error. Those models  provide a very conservative upper and lower bound for describing $F_\pi(Q^2>Q_{max}^2)$. The resulting uncertainty on the extracted pion transverse charge density is dominated by the incompleteness error at values of $b<$0.1~fm. The relative uncertainty in the region 0.1~fm~$<~b<~$~0.3~fm is smallest and the region above $b>$~0.3~fm is dominated by the need to truncate the Bessel series of Eq.~\ref{eq:theory:bigEq}.

A comparison of the pion to the proton transverse charge densities shows a larger density of the pion in the region $b<$0.2~fm. The two curves coalesce for values 0.3~fm~$<b<0.6$~fm, which may be interpreted in terms of the spatial structure of the proton consisting of a core occupying most of the volume and a meson cloud dominating only at large impact parameters. The coming together of the two curves at the edge of their density  suggest a common confinement mechanism for pions and protons. Future experiments at 12~GeV JLab and EIC will add high precision form factor data at higher $Q^2$ and would reduce the uncertainty, which in the region of $b>$~0.3~fm could be of great interest for studies of a common transverse charge density.

\vspace{0.6in}
\centerline{ACKNOWLEDGMENTS} 

We thank Pawel Nadel Turonski, Rolf Ent, and Adriana Rocha Lima for inspiring discussions and helpful comments and suggestions. This work was supported in part by NSF Grant Nos. PHY-1019521, PHY-1306227, and USDOE Grant No. DE-FG02-97ER-41014.


\end{document}